\documentclass[11pt,a4wide,twocolumn]{article}
\usepackage{amsmath,amssymb,enumerate}
\usepackage{amsfonts}
\usepackage{latexsym}
\usepackage{hyperref}

\catcode`\>=\active \def>{
\fontencoding{T1}\selectfont\symbol{62}\fontencoding{\encodingdefault}}
\newcommand{\nocomma}{}
\newcommand{\noplus}{}
\newcommand{\tmem}[1]{{\em #1\/}}
\newcommand{\tmop}[1]{\ensuremath{\operatorname{#1}}}

\newenvironment{enumeratealpha}{\begin{enumerate}[a{\textup{)}}] }{\end{enumerate}}
\newenvironment{enumeratenumeric}{\begin{enumerate}[1.] }{\end{enumerate}}

\newcommand{\tmfloatcontents}{}
\newlength{\tmfloatwidth}
\newcommand{\tmfloat}[5]{\renewcommand{\tmfloatcontents}{#4}
  \setlength{\tmfloatwidth}{\widthof{\tmfloatcontents}+1in}
  \ifthenelse{\equal{#2}{small}}
    {\ifthenelse{\lengthtest{\tmfloatwidth > \linewidth}}
      {\setlength{\tmfloatwidth}{\linewidth}}{}}
    {\setlength{\tmfloatwidth}{\linewidth}}
  \begin{minipage}[#1]{\tmfloatwidth}
    \begin{center}
      \tmfloatcontents
      \captionof{#3}{#5}
    \end{center}
  \end{minipage}}
\begin{document}
\onecolumn

\title{
  %\textbf{
  {\huge The conception of photons 
  %-- Part I
  }
  %}
  %\\
  %{\textsl{\large Planck, Einstein, and key events in the early history}}
}

\author{Urjit A. Yajnik}
\date{}
\maketitle

{\small {keywords} :
{Photons, light quanta, birth of the quantum, Planck spectrum, photoelectric effect, quantum 
indistinguishability, Bose-Einstein statistics, Glauber-Sudarshan representation, principle of linear superposition.}}

\tableofcontents

%\newpage
%\textbf{\large Part II}

\twocolumn
\begin{center}
{\huge 
%The conception of photons -- 
Part I}\\
  %}
  \medskip
  {\textsl{\large Planck, Einstein, and key events in the early history}}
%\begin{abstract}
\end{center}
 
 Abstract : In the year 1900 Max Planck was led by experimental observations to propose 
 a strange formula for the intensity as a function of frequency for 
 light emitted by a cavity made in a hot substance such as a metal. 
 %The formula reads
 %\[
  %\rho (\nu) = \frac{8 \pi h \nu^3}{c^3}  \frac{e^{- h \nu / k T}}{1 - e^{- h
  %\nu / k T}}
 %\]
%The usual rules of Thermodynamics according to Boltzmann would have led to something 
l%ike \( e^{-\mathrm{Energy}/kT} \). Thus an interpretation of the formula as describing 
%a gas of radiation contained in the hot cavity was problematic for two reasons : The 
%exponential in the  numerator could not be ascribed to energy of light according to known
%laws, and secondly the additional exponential in the denominator was not expected at all.
Planck provided a derivation based on peculiar properties to be obeyed by the emitters and 
absorbers in the cavity. I attempt to point out some nuts and bolts reasoning that could  have 
provided a clue to the physical reasoning.

In 1905, Einstein made the bold hypothesis that under certain circumstances, radiation could 
be absorbed and emitted as packets of energy and also propagated without spreading 
out like waves. Einstein was able to predict the formula for the photoelectric effect based on 
his hypothesis. While the formula was experimentally verified by 1913, his peers seem to have rejected 
its interpretation  in  terms of light quanta. Einstein himself was aware of its inherent contradictions. 
The first part of this article goes over this period of struggle with the photon concept, and sets 
the stage for the entry of S N Bose's critical contribution in 1923.

%It fell upon S.  N. Bose in distant East Bengal  in 1924 to  successfully derive the entire Planck 
%formula, including its enigmatic denominator, from  the assumption of light quanta treated as 
%a gas in the spirit of Boltzmann, but by incorporating a crucial new assumption about how they 
%fill up their available energy states. His communication of the result to Einstein, and its ready 
%acceptance, led to a flurry of new activities in Quantum Physics, becoming an integral part of 
%the final version of  Quantum Mechanics as we now know.
 
%\end{abstract}

%\twocolumn

\section
%{\tmtextsf
{Prologue }
%}

%\tmtextsf{\tmtextbf
\subsection{From an embarrassment to a paradigm shift}
%}

If ideas could speak they would tell us very strange tales. Firstly how the
idea arose is itself an interesting question. And it has any life of its own
only if the contemporary people accept it and propagate it as interesting.
There are many ideas for which it is said their time had come, and in this
case several different people independently think of the same thing within a
short period of time. Einstein's was an exceptional case to have come up with
not one but several different profound ideas within just a few years' time and
to have articulated them within a single year, 1905, and to have received a
quick acceptance for them. There is some evidence that Special Relativity was
being contemplated by several others, some of them stalwarts, at the same
time. But the conviction and clarity with which Einstein expounded its
fundamental nature probably got it instant fame.

However another of Einstein's profound ideas from the same year, contained in
a paper ``On a heuristic point of view concerning the production and
transformation of light'' had a rather different fate. This is the paper that
\ introduced the idea of what we now call a photon. It was the only paper to
take forward the rather heterodox ideas put forward by Max Planck about the
behaviour of light in his early papers in the year 1900. This too was written
with the same clarity and conviction that characterised Einstein's papers. Not
only was this idea not accepted, it was in fact considered to be an
embarrassment by his contemporaries. And even while he became reputed for his
other papers, which earned him a full professorship at Berlin, there was
pressure on him to withdraw this particular paper. In practical terms, the
patently ludicrous nature of the idea delayed his admission into the Prussian
Academy by several years, and therefore also probably delayed his Nobel prize.
It was not until S. N. Bose from India provided him with a ``missing link''
derivation in 1924 that Einstein himself fully, and \ the rest of the world
for the first time, became convinced of the correctness of the idea. It is
ironic indeed that Einstein received his Nobel Prize for none of his other
astounding ideas but precisely this one which was causing so much
embarrassment. However, the Prize was not given for the profoundness of its
core concept, but merely for it being a correct phenomenological prediction. \
And the award of the Prize, however circumspect, happened in 1921, before the
clinching proof provided by Bose. \

The goal of this article is to put these events in perspective, while also
discussing the science involved, along with the evidence it is based upon. We
begin with a discussion of what the challenge of Black Body radiation was,
followed by a possible reasoning that may make plausible Planck's path to the
leap into the unknown. Next I discuss Einstein's paper, his argument for the
core new concept it advanced and the possible reasons for the conviction
Einstein carried for an apparently irreconcilable stance which came to be
eschewed by all his colleagues. \ I also try to conjecture why it fell to S.
N. Bose in the distant colonial Indian university of Dhaka and with a gap of
two decades, to provide the systematic derivation. In the later \ parts of the
article I take up the aftermath of the revolution started by this paper, viz.,
the emergence of Quantum Mechanics. We shall be concerned mainly with \
Einstein's own response to the later developments, in that he came to
disapprove of the schema of the new Mechanics that his path breaking paper had
helped to unravel. I also briefly give the follow up story of the
circumstances that led to the development of the complete definitive theory of
light in the hands of Glauber and Sudarshan. The great debate of whether light
is a particle as Newton proposed, or a wave, as per Huygens' sophisticated
constructions, finds a culmination in this modern description. It does not
endorse either side as ``true'', but shows both descriptions as merely two
facets of a multifaceted, subtle phenomenon!

While the new Mechanics opened up the gates to new phenomena, new materials
and new forces of nature, its originator seems to have remained unconvinced of
its additional conceptual foundations. In this sense, this is the story of an
idea which was arduously protected by its proposer for decades under attack
from the contemporaries, but whose subsequent implications were rejected by
the same originator just after the idea received a resounding confirmation by
numerous experiments and an enthusiastic acceptance by the rest of the world.

\

%\tmtextsf{
\section{The antecedents}
%}

%\tmtextsf{
\subsection{The quiet before the storm}
%}
\label{subsec:quiet}

The end of the nineteenth century appeared to mark an epoch of triumphs in the
science of Physics. Heat, light and electricity, independent subjects under
Physics in any school textbook, were suddenly coming closer. The science of
Thermodynamics had been put on sound and consistent footing. Maxwell had put
together a comprehensive mathematical and conceptual framework of
Electromagnetism. The several laws due to Coulomb, Amp{\`e}re, Faraday and
others painstakingly put together over a century were beautifully united in a
common conceptual framework. An added wonder of this accomplishment was that
light for the first time could be clearly understood to be an Electromagnetic
phenomenon. Finally, radiated heat could be understood and a form of light.

It may be these developments that led Lord Kelvin to announce that all the
major discoveries of Physics had already been made and future explorations
will only help to improve the preciseness of the value of ``this or that
constant''. To be sure, Newton's schema of Mechanics was the conceptual and
mathematical framework to which all motion should conform. Euler,  Bernoulli,
Poisson and others had extended the Newtonian framework to deal with continuum
systems like fluids and solids. Heat had been recognised as a form of energy
and interconvertible with mechanical energy. Energy was also getting a place
in the scheme of Maxwell's Electromagnetism which in turn was also unifying
the phenomenon of light with the rest of Physics. These developments
comprehensively embraced all the domains of phenomena that Physics as a
mathematical science could embrace, placing them on a platform of universal
concepts and frameworks.

%\tmtextsf{
\subsection{The chinks in the armour}
%}

Yet, there were developments that needed reckoning, some happening soon after
Lord Kelvin's declaration. The biggest development that was to eventually
challenge all of Physics were brewing slowly  in Chemistry. Dalton's theory
of atoms and valence were very crucial developments towards atomistic and
unified theory of the world.  Combined with Avogadro's observation about a
standardised number of atoms that would occur in any material under specified
standard conditions, was putting an atomistic view of the world on a sound
footing.

A dichotomy immediately becomes apparent, since the continuum mechanics of
solids and fluids assumed substances to be ideal continua, while all the
materials were slowly being revealed as an agglomeration of only a fixed basic
list of ``atoms''.  To be sure, atomistic view of the world had long been
proposed in many schools of thought around the world. The reason is somewhat
obvious, with hindsight. Iron or copper from any mine in the world had the
same properties. Wood or oil always burned; water in any water body was more
or less the same substance, with some differences. Thus it was easy to guess
that there were some primary substances, with varying manifestations. Further,
since each had a characteristic property differentiating it from the others,
there had to be a basic unit that carried these special qualities. If all
materials were indefinitely divisible, it would be difficult to distinguish
between them in the ultimate limit of subdivision. On the other hand the
presence of a basic unit, the so called atom could be a candidate for
encoding the few basic properties like colour, smell taste etc that are
characteristic of the substance. Thus continuum mechanics could be
understood as a good effective description on a larger scale, while atomic
picture as the more fundamental one.

There was another direction where new phenomena demanded understanding. 
Kirchhoff and Bunsen had been carrying out experiments with metals heated to
very high temperatures. This was made possible by that humble equipment
available to every school laboratory today, the Bunsen burner. But the
efficiency with which it burned gas to produce complete combustion and a blue
flame made possible experiments that were previously difficult to perform. One
of the great observations of  Kirchhoff and Bunsen was that the spectra
observed in light from heavenly bodies like stars were exactly the same as
those that could be obtained from substances found on the Earth, such as
Hydrogen, Calcium, Sodium etc. We may consider this a next step in unification
since Newton. Newton had shown that curved paths of planetary motion were just
a generalisation of the fall of the apple on the earth. The heavenly bodies
obeyed exactly the same law as the bodies on Earth. Now Kirchhoff and Bunsen
were showing that the constitution of the substances in the distant stars was
the same as that of the substances on the Earth. Together with the atomistic
theory, this was showing that the Earth was just one among a large number of
bodies made up of the same materials obeying the same laws.

One more development that began to unfold in the 1890's was radioactivity.
Fluorescence was a known phenomenon and could be understood as an excited
state of an atom or a molecule. When Henri Becquerel first observed a
radioactive substance he mistook it for a more peculiar kind of fluorescent
substance. Henri Poincar{\'e} was the one to point out that the properties
that Becquerel had observed suggested that this peculiar radiation arose
spontaneously from deep inside the atom, and not merely from the excitation
and de-excitation of the same atom. As we now know, these were the first clues
to new fundamental force laws of nature, the Weak and the Strong nuclear
forces. The stupendously high energy outputs and the longevity of stars could
be understood only after  the nuclear forces were understood.

%\tmtextsf{
\section{The stage is set}
%}

%\tmtextsf{\tmtextbf{
\subsection{A challenge to theorists}\label{subsec:challenge}
%}}

Kirchhoff and Bunsen made a salient observation other than the universality of
substances mentioned above. It is a common observation that all metals when
hot begin to glow. They glow in a specific colour that is characteristic of
the metal that is being heated. However, Kirchhoff and Bunsen observed that as
the temperature becomes really high, the characteristics specific to the
substance grew gradually less important, and all the metals glowed in a
universal manner. They had made measurements of the intensity of the glow in
different wavelength ranges, i.e., at different colours of the spectrum.
Kirchhoff came to conclude that the rate of energy emission in a given
frequency range depends only on the temperature of the emitting metal, and
independent of the specific properties of the metal. As early as 1859
Kirchhoff wrote this as a paper challenging theorists to find an explanation.

In what follows we are not going to adhere to the original chain of
development of ideas. We shall adopt an ahistoric vain and pose ourselves
``what if'' questions such that if we could be selective about the sequence of
discoveries, what would be a good sequence in which to explore the facts so
that a coherent picture of the physical laws emerges inductively. The history
of the personalities and of the main events is nevertheless interesting and we
shall use that as the main framework for exposition, but in the interest of
keeping the emerging physical principles clear we may skip some false starts
and topical intermediate developments. In this vain we may proceed by
recapitulating Kirchhoff's observation as

\begin{enumeratealpha}
  \item the spectrum of emitted energy is independent of the substance
  emitting it, and
  
  \item the emitted total intensity depends only on temperature.
\end{enumeratealpha}
Considering the fact that individual substances are known to have their
characteristic frequencies in which they absorb or emit, we may interpret a)
to mean that the spectrum was somehow a property of light itself. Whereas b)
means that the only characteristic that could change from one situation to
another was the temperature. This in turn can be meant to read that light must
be subject to thermodynamic laws in a manner similar to atoms and molecules.

From the ordinary substances, standard gas laws were already known to exist.
Further, a microscopic picture existed from which to derive the bulk laws. The
distribution of kinetic energy among the atoms or molecules in an ideal gas
was known in the form of the Maxwell-Boltzmann formula,
\begin{equation}
  f (v) = \sqrt{\left( \frac{m}{2 \pi k T} \right)^3} 4 \pi v^2 e^{- m v^2 / 2
  k T} \label{eq:MaxBol}
\end{equation}
Here $v$ is a possible speed for the molecule, $m$ is the mass of a molecule,
$k$ is Boltzmann's constant appearing in all of Thermodynamics, and T is the
temperature expressed in Kelvin's absolute scale. The expression $m v^2 / 2$
is the kinetic energy of an atom or molecule. For the purpose of comparison
with what follows we can write down the kinetic energy density in an interval
of speeds $v$ to $v + d v$ as
\begin{equation}
  \rho_{_{M - B}} (v) = \frac{1}{2} m v^2 f (v) \label{eq:MaxBolDensity}
\end{equation}

In view of point a) of Kirchhoff's challenge, the effect needed \ to be
understood in its essentials, without interference from other incidental
effects. As is usually done in theoretical physics, to get to such a situation
some idealisation and simplification were introduced. It could be shown that
the property Kirchhoff was highlighting could be observed without other
encumbrance if one dug a small hole in the surface of a metal. Then the cavity
acted like a ``perfect emitter'', as well as a ``perfect absorber'', also
called a ``Black Body''. Kirchhoff's proposal of the frequency distribution
being completely determined by the temperature applied accurately to Black
Body radiation. The stated problem thus also came to be known as the Black
Body radiation problem. \

To keep the physics issues clear we now jump ahead here to the answer to this
challenge,  whose history will be dealt with later in Sec. 
\ref{subsec:desperation}. The distribution of electromagnetic energy into
frequencies as determined by experiment, and matching Kirchhoff's hypothesis
was discovered correctly by Max Planck in the year 1900. It looks like this
\begin{equation}
  \rho (\nu) = \frac{8 \pi h \nu^3}{c^3}  \frac{e^{- h \nu / k T}}{1 - e^{- h
  \nu / k T}}  \label{eq:Planck}
\end{equation}
Here $\nu$ (greek letter nu, distinct from the letter $v$ used above for
velocity) is the frequency of the light, $\rho$ (greek letter rho) is the
intensity of emitted radiation per unit volume per unit frequency interval,
$c$ is the speed of light and finally, $h$ is a new constant of nature, called
Planck's constant. We have written out the formula in the modern notation.
While Planck was quick to realise that he had identified a new constant of
nature, many aspects of the formula were going to remain rather unclear for
two decades.

Formulas (\ref{eq:MaxBolDensity}) and (\ref{eq:Planck}) are complicated even
for a college student. The main point to focus on is that the factor
\begin{equation}
  e^{- \tmop{Energy} / k T} \label{eq:BolFactor}
\end{equation}
occurring in formula (\ref{eq:MaxBolDensity}), an exponenetial, is known as
the Boltzmann factor, and a similar factor occurs in Eq. (\ref{eq:Planck}).
The Boltzmann factor is the intrinsic probability that a particle will be
found to have that particular value of energy in the hot medium at temperature
$T$. \ The product $k T$ which has the dimension of energy is determined from
the bulk measurement of temperature by say a thermometer, while ``energy'' in
the numerator is intrinsic energy of a microscopic member of the gas. Hence,
if we associated $h \nu$ with energy of light, then by analogy with formula
(\ref{eq:MaxBolDensity}), one could think of formula (\ref{eq:Planck}) as
relating to the ``gas of light'' or ``light gas''. In fact before Planck's
complete formula there was an ``almost'' correct formula of Wien, which read
\begin{equation}
  \rho_{_{\tmop{Wien}}} (\nu) = \frac{8 \pi h \nu^3}{c^3} e^{- h \nu / k T}
  \label{eq:Wien}
\end{equation}
Wien's formula had found considerable success in the short wavelength or high
frequency regime, but had begun to show deviation at long wavelengths and low
frequencies. But from theoretical point of view, it was a completely
unfathomable formula at that stage of knowledge. Neither the Boltzmann
exponential, nor the front factor of $\nu^3$ could be understood.

Based on Maxwell theory, light was known to be waves. The energy of a wave was
determined by its amplitude and not its frequency, i.e., by the extent of
vibration, not by the speed of vibration. Thus if a fundamental principle of
Thermodynamics called the ``equipartition of energy'' was applied, all the
modes of light would oscillate with the same amplitude to absorb the same
amounts of energy, in other words, no energy dependence in the form of the
Boltzmann factor would appear in the energy distribution function. This
observation was made by Lord Rayleigh in 1998 but seems to have escaped
attention. Further, Lord Rayleigh observed that from classical point of view,
a factor $\nu^2$ was to be expected for modes of oscillation  in three space
dimensions. He also noted that at sufficiently small frequencies, the
exponential in Wien's formula approaches unity, and then the classical
assumption of a distribution independent of frequency would become reliable. 
Based on this, he went on to propose that at low frequencies where the usual
thermodynamics begins to be applicable, the front factor should be $k T \nu^2$
instead of $\nu^3$. This is exactly what Rubens found two years later and
communicated to Planck prior to publication, and which spurred Planck to
arrive at the correct formula. Planck's 1900 papers however do not refer to
Lord Rayleigh's remarks which had appeared in the British journal
Philosophical Magazine in 1898. It is an interesting question whether Planck's
fortuitous foray could have been made two years earlier had he known of and
relied upon Lord Rayleigh's remark.  The question whether it was the language
barrier or a cultural difference in professional circles that kept the efforts
across the English channel sequestered from each other is also an interesting
one, because as we shall later argue, this may also have been the reason why
the discovery of the full law concerning photons awaited S. N. Bose for two
whole decades, and who also was far away from the flourishing centres of
science.

If we were very bold, (and willing to be ostracised by the community of
Physicists of that time), we could argue something like this. Since light was
known to be waves, this had to be a gas of waves, and Wien's formula should be
taken to mean that the energy of individual microscopic unit of this gas has
energy $h \nu$ as suggested by the Boltzmann factor, an association never made
before for waves. Further, combined with Lord Rayleigh's argument that the
front factor should have a $\nu^2$ just from the counting of modes, the
additional $\nu$ in the $\nu^3$ factor could be correctly understood as being
proportional to the energy. Now in analogy with $m v^2 / 2$ \ as the energy of
a molecule in formula (\ref{eq:MaxBolDensity}), and comparing Eq.
(\ref{eq:Wien}), we could begin to identify $h \nu$ in formula
(\ref{eq:Planck}) with the energy of light. Thus the spectrum would correctly
display the Boltzmann like distribution of energy. There would still be a
major discrepancy in the denominator, but we would be set substantially on the
right track.

As it turned out, the correct formula (\ref{eq:Planck}) was arrived at after
further experimental detail of the spectrum was known, and Planck would have
been greatly assisted in the process of trying to explain it, had he adopted
such a bold hypothesis. Planck however was a very conservative physicist, and
in any case, bold hypotheses without foundation are not the way of science. He
therefore first made sure the formula fitted the experimental data, and in a
later paper, to explain the origin of the formula, adopted what seemed like a
thermodynamic argument applied to absorbers and emitters in the walls of the
cavity. He did not focus attention on light itself, because that would have
led him immediately to the dead end suggested by Equipartition Principle
mentioned above. Planck's thermodynamic argument was circuitous and puzzling
to many physicists, but that was closest to anything like an explanation one
could advance with that state of knowledge.

Planck's caution in avoiding application of the thermodynamic argument to
light gas was justified. There was a fundamental difference in conception
between the two substances. Maxwell's theory relied on the fact that the
Electromagnetic field phenomena were continuum phenomena in the ideal sense.
There was no limit to how much you subdivide the space to be observed, there
would be newer degrees of freedom of the Electromagnetic field. This was
unlike other substances, where the atomic properties would become observable
beyond some extent of subdivision, and attempts to subdivide the space into
regions smaller than the atomic dimensions would not bring in any new degrees
of freedom.

Despite the difficulties of understanding formula (\ref{eq:Planck}), it was
good news. If we deviate from Planck and adopt the view that some kind of
thermodynamic argument could in fact be applied to light then thinking along
the lines of the bold hypothesis would be fruitful. Light had been understood
as a wave \ phenomenon, but here its collective system was cast in a form
similar to a gas of particles. Heat had been clearly understood as a form of
energy of molecular motion only half a century earlier, and the far infra red
light was understood as heat waves. Now it was being found that the energy
contained in light, when confined to a cavity, also had the properties of
heat, and had a temperature characterising its thermodynamics. Diverse
phenomena were coming closer and appearing to obey the same framework of laws.
We may think of this as the meta-principle that may have guided Einstein in
making the hypothesis even bolder than ours, to be discussed in Sec.
\ref{sec:LightQuantum}. However, a further mysterious difference was the
modified denominator of formula (\ref{eq:Planck}). In subsection
\ref{subsec:desperation}, we shall see how Planck could deduce the presence of
the extra $- e^{- h \nu / k T}$ in the denominator, but its full derivation
awaited S. N. Bose.

%\tmtextsf{\tmtextbf{
\subsection{Statistical Mechanics}
%}}

To further understand and appreciate  Kirchhoff's challenge we need a
diversion into Thermodynamics and its basis in Mechanics, as provided by
Statistical Mechanics.  Thermodynamics had made great strides, with the
recognition that heat was a form of energy and that pressure could be
understood as the collective average force exerted on the walls of the
container by the motion of myriads molecules. A major intellectual challenge
to full understanding of Thermodynamics as collective mechanical behaviour of
molecules was the fact that heat could only be dissipated, and if channelised
to produce usable mechanical energy, there were theoretical limits on what
fraction of it could be so converted.

A formula had been found by Nicolas Sadi Carnot, relating heat considered as
energy and its capacity to do work. This formula when examined showed that
there would always be some heat which will be wasted and cannot be converted
to usable work. Based on this fact, Rudolf Clausius developed the concept of
``entropy'' suggesting the sense of \ ``wasted'' form of energy ( trope
meaning turned away). For an amount of heat $\Delta Q$ being exchanged with a
heat bath at temperature $T$, the associated entropy \ $\Delta S$, quantifying
the irrecoverable part of the total energy, is given as
\begin{equation}
  \Delta S = \frac{\Delta Q}{T} \label{eq:entropy}
\end{equation}
Here $\Delta$, (upper case of Greek letter delta) is meant to suggest a small
change in, and a small amount of, the respective quantities $S$ or $Q$. This
profound and very subtle identification balanced the Thermodynamic equations
and accounted for all observations. However a microscopic explanation of
entropy at molecular level was lacking.

This gap in the understanding was sought to be fulfilled by Ludwig Boltzmann.
He introduced the concept of ``ensembles'', to mean the set of all possible
states the mechanical system could assume in principle. He then set about
proposing the rules of computation that would explain how the \ observed
Thermodynamics would emerge from this fundamental counting of all possible
configurations. He could obtain both the distribution law (\ref{eq:MaxBol}),
and the weightage factor (\ref{eq:BolFactor}) as a universal feature, and also
identify a fundamental quantity associated with ensembles as the entropy of
gases as quantified in eq. (\ref{eq:entropy}) from his fundamental postulates
of Statistical Mechanics. His famous formula reads
\begin{equation}
  S = k \log W \label{eq:BolEntropy}
\end{equation}
where $k$ as before is the Boltzmann constant, and $W$ is the number of all
the possible states that the system can attain consistent with general
conservation laws. In a sense then, the challenge thrown up by Kirchhoff's
observation was to identify and enumerate the list of possible microscopic
states of light treated as some kind of substance. Everybody took the Maxwell
theory as the basis of their computation and enumerated the fundamental
microscopic states accordingly. And the results were a disaster. They obtained
an answer that suggested indefinitely large contribution to energy at shorter
wavelengths\footnote{This is true for example if we take Lord Rayleigh's
proposal as described after Eq. (\ref{eq:Wien}) in subsection
\ref{subsec:challenge} to apply over the whole range of frequencies. If this
was taken literally, it would mean that the gas of light would have infinite
total energy. Lord Rayleigh perhaps anticipating this, had already proposed in
his paper appending the exponential suppression found empirically by Wien.}. 
This classical expectation  elaborated in 1904 by Jeans to counter Planck's
1900 hypothesis, came to be known as the ``ultraviolet catastrophe'' of Black
Body radiation. In a sense this indefinite growth in energy was to be
expected, since the electromagnetic field was assumed to be a continuum in the
Newtonian sense, consisting of newer degrees of freedom down to indefinitely
smaller scales of distance.

%\tmtextsf{
\subsection{An act of desperation}\label{subsec:desperation}
%}

We now return to the historical development. One person to face up to
Kirchhoff's challenge seriously was Max Planck. He was appointed a professor
at Berlin in 1889 and became full professor in 1892, to occupy the chair
previously occupied by Kirchhoff. He was well established but keen on making a
mark. He was the only theorist in this Department but it proved fortuitous for
him,  because in another laboratory PTR in Berlin (later to become PTB, the
German bureau of standards), Pringsheim and Lummer were working on
experimental determination of the spectrum, or the frequency dependence of
intensity of the Black Body radiation.  Ironically while the theoretical
spectrum failed in the short wavelength limit (newer degrees of freedom
emerging at ever smaller distances in the wave picture), Planck got a hint
about the correct answer from the knowledge of the long wavelength limit.

Interestingly, the spectrum of Black Body radiation was much more difficult to
measure at long wavelengths. At long wavelengths, Electromagnetic radiation is
essentially what we would call heat waves. These are very difficult to channel
and measure accurately. However Pringsheim and Lummer of Berlin had
painstakingly developed techniques that allowed them to measure the \ Black
Body radiation in the infra-red, ie., long wavelength region of the spectrum.
From Kirchhoff's time gatherings among professors' families were common in
Berlin, and Planck continued the tradition. One Sunday afternoon Rubens an
experimentalist visitor at PTR, was on a family visit to Planck's place for
tea. During this he revealed what he had begun to see, namely at long \
wavelengths, the spectrum was proportional to temperature, unlike at short
wavelengths. We can paraphrase it as per our discussion above to mean that at
low frequencies, the formula took the form
\begin{equation}
  \rho_{_{\tmop{Rubens}}} \propto \nu^2 k T \label{eq:RhoClassical}
\end{equation}
and no Boltzmann factor. As per the reasoning of Lord Rayleigh discussed below
Wien's formula (\ref{eq:Wien}), this relation was in keeping with Maxwell
theory as well as Equipartition Principle. One has to associate the same
energy with all the modes of the electromagnetic field as one would for
classical waves, and by equipartition theorem this would be $k T / 2$ for
every independent degree of freedom. The front factor $\nu^2$ is simply
related to the density of the possible modes at that frequency.

Thus one required a clever interpolation, matching Wien's enigmatic but
working formula Eq. (\ref{eq:Wien}) at high frequencies, and matching the
classical expectation (\ref{eq:RhoClassical}) at low frequencies. Based on
this hint, Planck could quickly work out a possible form of the correct
spectrum. Let us see how this could have been done. The exponential function
(\ref{eq:BolFactor}) which occurs in both formulas (\ref{eq:MaxBolDensity})
and (\ref{eq:Planck}) has an infinite but simple power series expansion
\begin{equation}
  e^{- x} = 1 - x + \frac{x^2}{2} - \frac{x^3}{6} \ldots
  \label{eq:expexpansion}
\end{equation}
Now if $x$ is a number small compared to $1$, say $0.1$, then $x^2$, $x^3$ etc
are much much smaller and can be ignored in a first approximation. So we see
that if we consider the expression $1 - e^{- h \nu / k T}$, then when $h \nu /
k T$ is small, it is approximated by just $h \nu / k T$. Since this appears in
the denominator of (\ref{eq:Planck}), we get $\nu^2 T$ instead of $\nu^3$ in
the numerator of the formula\footnote{This way of thinking looks even easier
if the formula (\ref{eq:Planck}), aside from the front quantities, is written
in the form $1 / (e^{h \nu / k T} - 1)$. Equivalently if one looks for
temperature to enter only through Boltzmann factors placed suitably, this is
the simplest placement to obtain the required long wavelength behaviour.}.
This happens for small values of $\nu$, i.e., for large values of the
wavelength $\lambda = c / \nu$. This linear dependence on temperature was
exactly what Rubens was reporting to Planck, and which was being confirmed by
others such as Kurlsbaum, Pringsheim and Lummer.

Having confirmed the correctness of the spectrum he was predicting, Planck
would have set about working out a physical explanation or pinpointing the
assumptions that would underlie such an answer. Note that the law for the
molecules (\ref{eq:MaxBol}) has no such $- e^{- h \nu / k T}$ in the
denominator. Such a modification could also not come from any accidental
specific properties of light. It would have to emerge through some reasoning
to be on the same footing as the fundamental weightage factor
(\ref{eq:BolFactor}) of Boltzmann. Try as he would he could not find a
convincing reason. \ In later recollections Planck referred to his avid
attempts to prove the formula as ``an act of desperation'' (or a desperate
attempt, sometimes also translated as an act of despair by the fact that none
of the known theories worked). And he did somehow arrive at an explanation.

We now attempt a ``back of the envelope'' or nuts and bolts reasoning for how 
he may have arrived at such an explanation. Let us now use another fact of 
algebra : when a quantity $y$ is
small, (actually just less than 1 in magnitude is enough) we have the formula
for the sum of the infinite series
\begin{equation}
  1 + y + y^2 + y^3 \noplus + \ldots = \frac{1}{1 - y} \label{eq:geomseries}
\end{equation}
Now $e^{- h \nu / k T}$ is always less than $1$ if $h \nu / k T$ is positive,
and which it is on the physical grounds of positivity of $h \nu$ interpreted
as energy. So \ the required condition for the expansion (\ref{eq:geomseries})
to be applicable is satisfied. We rearrange the second fraction on the right
hand side of eq. (\ref{eq:Planck}) to read
\begin{eqnarray}
 &{\phantom{1}}& \frac{e^{- h \nu / k T}}{1 - e^{- h \nu / k T}}\nonumber\\ 
 & = & e^{- h \nu / k T} (1 +  e^{- h \nu / k T} + e^{- 2 h \nu / k T}\nonumber\\ 
  && + e^{- 3 h \nu / k T} + \ldots)  \nonumber\\
  & = & e^{- h \nu / k T} + e^{- 2 h \nu / k T} + e^{- 3 h \nu / k T} \nonumber\\ 
  &&+ e^{- 4 h \nu / k T} \ldots  
    \label{eq:PlanckGeomSeries}
\end{eqnarray}
Here we used the rules about exponentials viz., for any integer $n$, $(e^y)^n
= e^{n y}$. We now see a summation of Boltzmann factors of the type of eq.
(\ref{eq:BolFactor}). Planck may well have set about giving a microscopic
argument for this particular summation. \ Planck used the concept of entropy
and Boltzmann's method to derive his answer. In doing so however, it became
necessary for him to ascribe some peculiar properties to the emissions and
absorptions by atoms and molecules in the walls of the cavity treated as
oscillators in interaction with the radiation. Considering that Eq.
(\ref{eq:PlanckGeomSeries}) contains a string of Boltzmann factors containing
integer multiples of a basic frequency $\nu$, he was led to the following
hypothesis. This was that an oscillator of frequency $\nu \nocomma$, could
emit or absorb radiation only in integer multiples of the basic unit $h \nu$.
There was no reason in classical radiation theory for the absorptions and
emissions to be ``quantised'' in integer units, nor was the energy \ of
oscillations ever found to be proportional to frequency $\nu$. Of course, nor
had anyone previously encountered a physical constant $h$ of the given
magnitude and dimensions. 

What Planck needed to assume to make sense of this formula was very
startling. Given an oscillator of frequency $\nu$, it could absorb and emit
energy only in integer multiples of the basic energy unit $h \nu$. This was
the only way the string of Boltzmann factors in the summation made sense. In
arriving at this explanation he came very close to the explanation S. N. Bose
finally provided to his formula. In the present section let us only note this
particular assumption as implemented by Planck, \ viz., given a total number
say $N$ of ``energy units'' $h \nu$, one should consider all possible
assignments of occupation numbers $n_1$, $n_2$, to the various excited levels,
such that \ $n_1 + n_2 + \ldots = N$. This is the crucial assumption. Then by
Boltzmann's methods one could prove that when one asks for the maximum entropy
configuration among such assignments, letting $N$ become large, one
automatically comes to the answer eq. (\ref{eq:Planck}), implicitly, the
summation of eq. (\ref{eq:PlanckGeomSeries}).

%\tmtextsf{
\section{The light quantum proposal}\label{sec:LightQuantum}

\subsection{A heuristic viewpoint }\label{subsec:heuristic}
%}

Einstein saw the problems with understanding Planck's formula to be of a
fundamental nature. He begins in his paper by pointing out that a profound
difference exists in our conception of material particles on the one hand and
radiation on the other. Both carry energy and several common attributes, but
radiation was presumed to possess physical degrees of freedom down to
infinitesimally small length scales, while material particles being discrete
did not possess any new physical degrees of freedom smaller than their size.
According to him the solution resided in giving up the simplistic view of
radiation as continuum. While all the macroscopic phenomena connected with
radiation like reflection and refraction were well explained by the continuum
wave character, he believed this behaviour of light did not persist under all
circumstances. Particularly, he noted, the usual phenomena occur under
conditions of high intensity and long wavelengths. He noted three new
experimental results that had emerged in the late 1800's which also seemed to
be demanding an explanation. These were photoluminescence, photoelectric
effect and photoionisation. He seized on the possibility that the problems
faced in understanding ``light gas'' had not so much to do with the gaseous
phase, the presence of the cavity and so on, but to do with some fundamental
properties of radiation itself, which were shared in situations other than in
gaseous phase. Thus Einstein begins with a bold announcement in the
introductory section of his paper,

\begin{quotation}
 ``According to the assumption considered here, in the propagation of a
    light ray emitted from a point source, the energy is not distributed
    continuously over ever increasing volumes of space, but consists of finite
    number of energy quanta localised at points of space that move without
    dividing, and can be absorbed or generated only as complete units.''
\end{quotation}

It is a revolutionary idea when we think about it even today. Electromagnetic
theory works so well in explaining all the engineering phenomena based on the
assumption that the fields are a continuum. Did just the one or two new facts
demand so radical a change in thinking to lead one to say ``... energy is not
distributed continuously over ever increasing volumes of space, but consists
of finite number of energy quanta ...''?

The subject of theoretical physics is about achieving simplicity and economy
of thought. It hinges on identifying core concepts and their relationship to
various possible environments in which they occur. The challenge is to
identify these core concepts in a way that they remain  applicable to all
phenomena and also in a way that their relationship to all possible
environments is  of a universal nature. Here by ``environment'' we mean
processes such as emission and absorption where the radiation encounters other
systems or gets transformed. What Einstein was pointing out was that at least
three phenomena, photoluminescence, photoelectric effect and cavity radiation 
could not be reconciled with the conception of light as a continuum wave
phenomenon. Of course if one were so bold as to propose an alternative
conception, there was a need to reconcile it with the standard one under
standard situations. Here Einstein faced a difficulty. He did not know how the
wave picture would yield to his new bold particulate picture in terms of
``energy quanta'' whose ``energy is not distributed continuously over ever
increasing volumes of space''. But his stipulations for the new concept are
laid out with legal precision in the quoted paragraph. Indeed Einstein had to
face this enigma of irreconcilability for two more decades.

However, Einstein could have had his authority drawn from  another, even broader
consideration, viz., the need to reconcile the corpuscular conception of
matter with the continuum conception of radiation\footnote{It is reported that
Einstein was fascinated both by Maxwell's grand synthesis as well as
by corpuscular conception of matter. Interestingly, Avogadro's hypothesis
regarding the universality of the number of corpuscles in a gas under standard
conditions had taken a whole century to be accepted. It had a crucial role in
Boltzmann's theses on microscopic origins of Thermodynamics. Einstein's 1902
doctoral thesis concerned an experimental determination of Avogadro Number. }. 
His stance seems to suggest that the unity of core concepts for matter and radiation 
was more important to him than reconciling the two alternative descriptions of radiation. 
This could be the vision that gave him strength to hold on to his 
rather radical concept.

%\tmtextsf{
\subsection{Entropy as a hint to discretisation}
%}

We now proceed to give the original argument according to Einstein why the
Planck formula must be read as a formula applied to Thermodynamics of ``quanta
of light''. Here the notion of quanta, or or equivalently, discreteness enters
because the energy of radiation with frequency $\nu$ turn out to be integer
multiple of the basic unit $h \nu .$ \ As we shall explain in the next
subsection \ref{subsec:photoelectric}, we find that Einstein uses the
Statistical Mechanics definition of entropy and its relationship to the number
of gas molecules, but never the detailed derivation along the lines of
Boltzmann. Accordingly, Einstein first observes that the Planck formula and
its ultraviolet limit known as the Wien law give a relation between entropy
$S$ and volume $V$ of the subsystem of the radiation of frequency $\nu$ with
energy $E_{\nu}$ as
\begin{equation}
  S - S_0 = \frac{E_{\nu}}{\beta \nu} \ln \left( \frac{V}{V_0} \right)
  \label{eq:EinEntWien}
\end{equation}
where the quantities $S_0$, $V_0$ are the values at some convenient reference
point and $\beta$ is a constant. Next he notes that according to Boltzmann's
definition of entropy, the entropy depends on the number of particles $n$ with
a reference value $N$, and the volume occupied, according to the formula
\begin{equation}
  S - S_0 = R \left( \frac{n}{N} \right) \ln \left( \frac{V}{V_0} \right) =
  \left( \frac{R}{N} \right) \ln \left( \frac{V}{V_0} \right)^n
  \label{eq:EinEntBol}
\end{equation}
where $R$ is the gas constant appearing ideal gas equation of state. Thus,
using the rules of logarithm, if we rewrite (\ref{eq:EinEntWien}) as
\begin{equation}
  S - S_0 = \frac{R}{N} \ln \left( \frac{V}{V_0} \right)^{(N E_{\nu} / R \beta
  \nu)} \label{eq:EinEnt2Bol}
\end{equation}
then we have the correspondence that there is a number $n$ associated with
radiation of energy $E_{\nu}$,
\begin{equation}
  n \equiv \left( \frac{N E_{\nu}}{R \beta \nu} \right) \label{eq:EinQuant}
\end{equation}
so that $n$ can be identified with ``the number of quanta'', a concept which
is devoid of meaning in wave theory of Maxwell. Needless to say the
phenomenological constants in equation (\ref{eq:EinQuant}), which is written
here in its historical form, actually work out to just reproduce the Planck
constant $h$. In other words, eq. (\ref{eq:EinQuant}) is nothing but the
famous law $E_{\nu} = n h \nu$ with $n$ a natural number. We may think of
making such an association as in eq. (\ref{eq:EinQuant}) also as a theorist's
opportunism. Armed with this interpretation of Planck formula, Einstein
proceeds to apply the concept of quanta to the then yet unexplained results of
some controlled experiments.

It is worth emphasising at this point that Einstein deduced the photon
hypothesis through Wien's formula. He did not make use of the full Planck
formula, and for that enigmatic formula he advanced no explanation. He merely
proceeded by conviction to make further predictions based on his conception of
photons. In hindsight we know that it is the additional $e^{- h \nu / k T}$ in
the denominator which is the characteristic of a quantum gas. Ironically this
was not the hint that led him to the quantum proposal. His reasoning was more
along the lines of our ``bold hypothesis'', discussed in sec.
\ref{subsec:challenge}. This was surely revolutionary in proposing packets of
energy of value proportional to frequency and propagating undivided in a
specific direction from source to receiver, where Maxwell theory assumed
waves. Yet, the real bombshell of how uncannily different quanta are from
classical conception of particles was not to arrive until S. N. Bose's 1924
derivation would be further dissected.

%\tmtextsf{
\subsection{Photoelectric effect}\label{subsec:photoelectric}
%}

Sparkling and colourful objects like prisms, precious stones and so on were
one impetus to research on light. The end of the nineteenth century saw
another class of phenomena, glowing substances, that attracted researchers'
attention. It may be noted that often such research was not as restrictively
channelled as a ``Physics'' experiments but rather a part of study of natural
environment, with the aim of collating, classifying and archiving along the
lines one would do with rocks, minerals, or even birds, plants and insects.
Geophysical exploration in fact appears to have been a big hobby enterprise of
people of leisure, with there being learned societies waiting to hear eagerly
of new discoveries from ever newer habitats and unexplored corners of the
world. This was within the general ethos of nineteenth century Europe. We
may mention in passing that the discovery of radioactivity by Henri Becquerel
was somewhat of an accidental discovery made in the course of such
explorations.

In this list of things being explored was photoluminescence. There were
phosphorescent and fluorescence substances that after being exposed to
sunlight, seemed to retain the energy and continued to glow in the absence of
the Sun. Such glow eventually faded within a few days depending on the
substance. Despite the variety of substances, the phenomenon seemed to be
governed by Stokes law which stated that the frequency of light emitted was
always less than the largest frequency within the absorbed light. In classical
electromagnetism, it was always the amplitude of light, i.e., the extent of
undulations of the fields (actually the square of the latter, viz., the
intensity) that was a characteristic of the energy the light wave carried.
However, according to the Planck relation $E = h \nu$ it is the frequency of
undulations that determines the energy. Thus the photoluminescence law could
not be suspected to be essentially an energy conservation law in classical
theory. But from the Planck relation one can see that the limitation on
frequency is the statement that emitted light cannot be of greater energy than
the absorbed light. This law would then make common sense, because if a
substance began to emit more energy than it absorbed it could not remain a
stable substance for long. On the other hand the lower energy emissions were
only dissipating the energy stored during the day, with the substance
returning to its normal stable state after the emissions. 

For Einstein, who had grasped the generality of the law $E = h \nu$ as valid
intrinsic relation for light quanta, the above logic must have been immediate.
But the world was still believing along with Planck that his relation somehow
came into force only for the energy exchange between the ``oscillators'' in
the walls of a cavity and the radiation trapped within the cavity. They had
not associated it with a property of light itself.

The law regarding photoluminescence was already an empirically observed fact,
and Einstein gave an explanation for it. The more radical proposal to be made
by Einstein was regarding photoelectric effect. Here where the data were still
somewhat unclear, he made a radical hypothesis based on some general clues,
and then proposed a specific law, to be verified by experiments. As mentioned
earlier, Einstein had come to the realisation that the light quantum or photon
picture gained validity in the limit of very low intensity but short
wavelengths. In the case of photoelectric effect, the salient features to
emerge were again of a cut off value of energy and a linear dependence of a
measurable quantity on frequency.

In photoelectric effect, impinging radiation was found to eject electrons
from alkali metals. Using vacuum tube techniques and electrostatic plates, it
was possible to channel the ejected electrons and observe them as current. The
first salient feature to be discovered by Heinrich Hertz and by Hallwachs was
that there seemed to be a maximum value to the kinetic energy of the ejected
electrons. There was a ``stopping potential'' which could be applied to the
electrons, which would balance the maximum kinetic energy of the electrons.
Further, this maximum kinetic energy seemed to depend on frequency of the
radiation. Increasing the intensity of the radiation did not impart greater
kinetic energy to individual electrons. The situation has an analogy to the
photoluminescence case, if we remember that energy balance of individual
emission processes is determined by frequency and not by intensity. Finally,
if the stopping potential was made large enough, no electrons would reach the
anode gathering the current. The effect was first discovered by Heinrich Hertz
in 1880's followed by investigations in the case of alkali metals by Hallwachs
and later around 1903, the effect studied by Hungarian physicist Philipp
Lenard for gases which underwent ionisation. Lenard also reported a dependence
of the energy of ejected electrons on the frequency rather than intensity of
the radiation. 

Putting these things together, while they were still considered somewhat
controversial, Einstein advanced a clear cut equation for the maximum kinetic
energy of the ejected electrons,
\begin{equation}
  \tmop{KE}_{m a x} = h \nu - W \label{eq:Photoelectric}
\end{equation}
Here the quantity $W$ is the largest value of the stopping potential that
would extinguish the current, and it depended on the specific substance under
study. However, the first term was universal, independent of the system being
investigated, whether metal or gas.  It depended only on the frequency of the
radiation being sent in and the new constant of nature, $h$. As in the case of
photoluminescence, if you are tuned to the idea of packets of energy, and
frequency as the determinant of their energy, then this is simply an equation
of energy conservation. $W$ is simply some threshold energy the electron must
acquire, by deducting it from the impinging energy $h \nu$, to come free of
the substance from which it is being ejected.

This concludes the first part of our journey into the origins of the uncanny idea 
of photons.  In the next and concluding part we shall see how it contained the 
seeds of Quantm Mechanics, and the path from here that led to the full quantum 
theory of light.

\newpage
%\onecolumn
\begin{center}
 {\huge The conception of photons -- Part II}\\
  %}
  
  \medskip
  {\textsl{\large Bose's derivation, and the complete quantum description of light}}
\end{center}
\medskip
%\textbf{\large Part II}
%\twocolumn
Abstract : In this second part of the article we present how S N Bose's 1924 paper 
provided a systematic derivation of Planck formula using the conception of photons,
filling the major lacuna that was preventing the acceptance of the photon concept 
by the  Physics community. This derivation further widened the chasm between
classical conceptions and the actual behaviour of the microscopic world as already 
heralded by the photon proposal. In particular, the very concept of a quantum as an 
``independent'' entity even when not interacting with other entities is rendered 
invalid. Classical intuition was subverted in Bose's derivation by a new rule, regarding 
counting of independent states of the system rather  than counting individual quanta. 
We discuss the implications of the Quantum Mechanics that
eventually emerged, showing that the seeds of some of its uncanny conceptual content were
already foreshadowed in Einstein's proposal. While he was instrumental in setting 
off the revolution, the full implications of the revolution became unpalatable to him.
We may expect that as experiments make the quantum world more familiar, the currently
projected enigmas will gradually disappear.

%\twocolumn
\section{Towards the birth of the quantum}
%\tmtextsf{
\subsection{Seeds of the dreaded rules of the Quantum?}
%}
In the first part we saw how Einstein arrived at the famous formula applicable
to photoelectric effect,
\begin{equation*}
  \tmop{KE}_{m a x} = h \nu - W 
  %\label{eq:Photoelectric-again}
\end{equation*}
According to him, once light in this setting was understood to behave like packets of 
energy, the above formula was simply energy conservation formula. The proposal that
the energy of the light quantum should be proportional to its frequency ran against
the grain of Maxwell's electromagnetism where light could be shown to be a phenomenon
similar to waves in any medium.

But the surprise of this proposal is not restricted to this little paradox. The import 
of the ideas we have now covered -- and presumably accepted by you dear reader, as 
valid, -- is truly stupendous. Sometimes one wonders whether
Einstein fully grasped the extent of damage his proposal was doing to some of
the well established classical notions. Let us assume as clearly articulated
by him, that the emitted light was going to proceed without spreading out, as
an undivided packet of energy into a specific direction. Considering that the
emitting body was a point like object such as an atom or a molecule, one is
immediately faced with the question, ``which'' direction will the emitted
quantum proceed in? Even if the emitting body had a size, it could well be
very simple, say the Hydrogen atom, which can be presumed to be spherically
symmetric. Then simple classical reasoning would suggest that the radiation should
emerge as a spherical wave respecting the symmetry of the emitter. But according to
the photon hypothesis, the emission process must choose a preferred direction of emission.
%randomly, conflicting with the intrinsic symmetry of the "shape" of the emitter. 
What fundamental principles govern the choice of this direction are not spelt out by the
new stipulation. Yet, with a century of experience of Quantum Mechanics we know that 
this is indeed how the emission of a photon occurs and in a sense typifies the nature of all 
quantum processes. Only under repeated identical observations can
we establish the overall isotropy of the emission phenomenon, while in an
individual event, the symmetry will not be respected. We may also cite another example
of the often studied ``particle in a box''. Consider an electron confined within a box but 
with no other interactions. From quantum mechanics we find that its location is not evenly 
distributed within the box. Depending upon  the state it is in, it will be found preferentially 
at selected locations,   violating the homogeneity of the container. However observations 
of a large number  of such boxes will indeed restore the homogeneity of locations.
To repeat, the earliest hypothesis made by Einstein already encodes the principle now used 
by all practitioners of quantum mechanics, viz., isolated quantum processes have to occur 
with one specific eigenvalue of the concerned observable revealed in a given experiment.

Thus, in a sense, the seeds of the dreaded Quantum Mechanics were already sown
in Einstein's original proposal when he generalised Planck's law originally proposed for
a radiation gas to individual events of emission and absorption. But an
equally drastic phenomenon of nature had not yet been articulated, and it
awaited the correct Boltzmann ensemble of photons as conceived by S. N. Bose
two decades later. And this phenomenon is the intrinsic indistinguishability
of quanta which makes us realise that quanta are not at all ``particles'' such
as billiard balls we are familiar with, but profoundly novel entities.

%\tmtextsf{
\subsection{Opportunism of theorists}
%}

We may view the bold attempts of both Planck and Einstein as an opportunism of sorts,
the readiness to jump into the unknown, abandoning the comfortable territory,
for the possibility of obtaining a correct answer. As we noted earlier, Planck
later came to consider his \ effort as ``an act of desperation''.

Einstein on the other hand, faced a stigma. While he became famous for his
Special Relativity, the famous relationship between rest mass and energy etc.,
he was under pressure from senior colleagues to retract his radical ideas
about discrete nature of electromagnetic phenomena. Specifically, it was
creating difficulty in getting him elevated as a fellow of the Prussian
Academy. Despite being nominated several times, the committee examining his
case seemed to choke at this particular paper. It is reported that in a
subsequent nomination his proponents even attempted an apology on his behalf,
something to the effect that occasionally in his eagerness to explicate very
difficult phenomena he is led to rather radical proposals, but this need not
be held against him etc.

After this went on for several years, he was forced to issue some public
clarification about his rather radical and unsavoury paper at the Solvay
conference in 1911. But he stood up to the stalwarts, asserting that ``I must
insist on the validity of the new concept at least within the domain of
phenomena for which it furnishes an explanation.'' His statement made in
German was however was so carefully worded that it was interpreted as him
having reservations about this concept, at least in the English speaking
world. In fact Robert Millikan who confirmed the photoelectric effect experimentally, 
in his 1916 paper refers to EinsteinÕs quantum proposal as ``bold, not to say reckless'', 
considers it to have been ``generally abandoned'', and in his conclusions states that 
the proposal ``... is found so untenable that Einstein himself, I believe, no longer holds to it.''

By 1908 Einstein became preoccupied with General Theory of Relativity and seems
to have not returned to the question of light quanta.
It is notable that he also proposed an explanation for the behaviour of
specific heats of solids based on quantum vibrations in 1907 which met again
instant success as a general idea though not quite correct in detail. But his
photoelectric effect explanation was shunned by all experts. This caused
difficulty in his becoming a member of the Prussian Academy, and since
membership of national academy is a natural step towards the Nobel Prize, also
a delay in his getting that coveted Prize. Einstein was quite a celebrity
based on his Special Relativity and was becoming the next genius after Isaac
Newton with his formulation in 1915 of the General Theory of Relativity. But
the stupendous intellectual achievement of Special Relativity did not meet the
criteria of new phenomenological content required by the Nobel committee,
while the discovery of phenomena that would conclusively establish General Relativity
remained far in the future.

In 1914 Robert Millikan confirmed the formula proposed by Einstein, yet
nobody including Millikan seemed to believe the conceptual basis of the
formula. Thus it was that with much struggle the well wishers of Einstein and
no doubt well wishers of the subject of Physics managed to convince the Nobel
committee to award the 1921 Prize to Einstein for his discovery of the
photoelectric formula. And thus the Prize was awarded to him, taking care to
state in the citation that it was "for his services to Theoretical Physics,
and especially for his discovery of the law of the photoelectric effect". Note
that it is not for the correct \ conceptual basis or theoretical explanation
of the effect, it is merely for the correct ``discovery'' of the law, the
prediction of the equation verified by Millikan.

While talking about opportunism, let us jump ahead a little and refer to sec.
\ref{sec:StatesandQuanta} where we discuss the core new concept underlying 
the contribution of S. N. Bose. It is common to note in critical assessments that
this derivation is technical, brief and while it proves the formula, does not
sufficiently explicate the new assumptions involved. One has to note however
that while Einstein was bold enough to move ahead to making a new prediction,
he made no attempt to explain the additional $- e^{- h \nu / k T}$ term in the
denominator Eq. (\ref{eq:Planck}). While Planck gleaned the formula, and Einstein could grasp the
quantum nature of the phenomenon, it was Bose who for the first time clearly
derived the whole expression from Boltzmann's ensembles, also incorporating a
revolutionary counting for photons.

\subsection{A curious case of inadequate diffusion of scientific knowledge?}

It is important at this point to note a few ironical quirks of history and
personalities. Einstein in this paper of 1905 is somewhat circumspect of the
methods used by Boltzmann. He does use the microscopic picture of entropy as
proposed by Boltzmann. But he carefully avoids using the method of ensembles.
For one he seems to think that listing all the members of an ensemble -- all
the possible states a system can possibly attain consistent with energy
conservation -- is still no guarantee that one has enumerated all possible
dynamical effects which occur when a system is undergoing time evolution. This
scepticism is similar to what has come to be called the question of
ergodicity, but the way Einstein states his objection it seems to be even
stronger than the question of ergodicity.  Secondly we are told by the editor
John Stachel of the collection of Einstein's papers during that ``miraculous
year'', that there were more reasons for which Boltzmann's contemporaries did not agree
with him in detail though many agreed in principle. And this was because
Boltzmann had a verbose writing style and the definitions of the concepts he
would propose were not sharply defined and would seem to change even within
the course of the same long essay. As we now know there were also a few errors
of normalisation in his formulae.

All the issues associated with Statistical Mechanics had been adequately
addressed by Josiah Willard Gibbs in the USA by the turn of the 1900's. Had
Einstein accessed that treatise, his doubts would probably have been addressed
and he would have proceeded to give a detailed derivation of the Planck
formula starting from his fundamental conception of photons, using the
techniques of Boltzmann, the same way the latter had provided microscopic
explanation of classical Thermodynamics. But this was not to be. Probably
because Einstein did not read English back then and also perhaps because the
centre of gravity of science and intellectual discourse was Europe and Gibbs's
treatise was slow in gaining acceptance there. We may then summarise the
impasse in the progress towards full understanding of the Planck formula on
two ironical circumstances : Albert Einstein firmly believed in photons but
would not produce a proof using the Statistical Mechanics, while the rest of
the world refused to believe in photons but certainly had many experts who
knew the latest reliable methods in Statistical Mechanics but who probably did
not bother to apply them to a gas of photons.

%\tmtextsf{
\subsection{Confirmation from far away, far later}
%}

It thus fell upon Satyendra Nath Bose, a professor in Dhaka (or Dakka)
University in 1924 to produce the required proof. Bose as a younger man
venerated Albert Einstein, and being far away from the European crucible of
science was perhaps immune to the prejudice prevailing against the notion of
photons. Further, as a brilliant scholar he had no doubt mastered the methods
of Statistical Mechanics, again without too much prejudice because being from
colonial India he had ease of access to the English source material, probably
including Gibb's treatise. Thus it was that he set about ascribing a
discretised character to the phase space of radiation. In Mechanics, where
both positions and momenta of particles need to be considered as independent
variables, the word phase space refers to the abstract space labelled by this
combined set of coordinates. He made the assumption that in line with quantum
principles, the phase space needs to be divided into discrete cells of size
$h^3$, a quantity whose unit dimensions match those of a volume element in
phase space. To this author's knowledge this was also the first calculation of
density of states for quantised bosons. The previous calculations had
introduced frequency dependent volume factors in phase space within the wave
picture. Bose's partitioning of the phase space is what we now call box
normalisation, and it is clear from his paper that this was very important
conceptually to Bose as the full package of the quantum hypothesis.

He then proceeded to list the possible states of the ensemble of photons and
inadvertently distributed the photons in available phase space boxes without
any discrimination among them. He then applied Boltzmann's method to identify
the equilibrium distribution which would dominate. It yielded exactly the
formula due to Planck.

It is not possible to go into the details of S. N. Bose's all too brief but
paradigm setting paper. But he had the full answer. There was some imagination
and then there was precise logic and a computation. Neither desperate nor
opportunistic, this derivation had the entire formula of Planck proved from
first principles of Statistical Mechanics and the conception of radiation as
photons. It is said that he sent his paper in English first to the Philosophical 
Magazine in 1923. But  it was rejected. He then sent his paper to Einstein 
addressing him as Respected Master.
Einstein immediately grasped the significance of this paper. This was the
calculation he had sought for the previous two whole decades. He translated it
and communicated it to the Zeitschrift f\"ur Physik. He then proceeded as a follow
up to work out the consequences of the new method of calculation applied to
massive particles. The combined general formulation is called 
Bose-Einstein Statistics to distinguish it from the classical Maxwell-Boltzmann
Statistics.

Bose's one off contribution has evoked puzzlement and its far reaching
implications also perhaps jealousy. True, he did not have a consistent output
of scientific contributions within that subject area like a European
scientist. Being sensitive to the condition of his country he shifted his
attention to practical problems of semiconductor devices. But nobody denies
the brilliance and scholarship of Bose. It appears that he himself did not
grasp the novel assumption he had inadvertently made in his derivation. While
the discrete partitioning of the phase space is an important step, the success
of the derivation relies on an additional crucial assumption. If we read the
wording of how Planck finally convinced himself of his derivation in the year 1900,
we see a parallel. Planck was thinking of energy as a generic quantity to be
distributed among those oscillators in the walls of the cavity. And he spoke
of distributing ``energy units'' into the available excited states of the
oscillators. Of course with hindsight we know the oscillators were a
completely unnecessary scaffolding. Bose on the other hand had to contend with
the same energy units, now conceived as photons, themselves the objects of
Statistical Mechanics to be handled by set rules. The scaffolding of cavity
oscillators was abandoned once and for all. And he implicitly distributed
photons among their own available energy levels according to the same
indistinguishability approach as Planck. As we elaborate below, this is the
key novel assumption, which naturally produces the denominator of Eq. (\ref{eq:Planck}) 
of part I, viz.,
\begin{equation*}
  \rho (\nu) = \frac{8 \pi h \nu^3}{c^3}  \frac{e^{- h \nu / k T}}{1 - e^{- h
  \nu / k T}}  
  %\label{eq:Planck-again}
\end{equation*}
%(\ref{eq:Planck}) 
without any reference to any oscillators. Bose himself
missed this particular fact, and there is nothing to indicate that even
Einstein understood it at the time of communicating his paper. It was indeed
something very very subtle. Thus to Bose we may attribute the credit for arriving 
at the ``light gas'' distribution formula by applying the principles of statistical 
mechanics directly to photons considered as fundamental entities.

%\tmtextsf{
\section{Quantum Mechanics}
%}

%\tmtextsf{
\subsection{Ideas whose time had come}
%}

Between 1905 and 1924 Einstein returned to the physics of light a few times,
but the issue of validation of the photon hypothesis remained unresolved. In 1917 
as the general relativity revolution was catching on, he devoted attention to radiation 
again, and wrote his famous insightful paper on the so called A and B coefficients, 
concerning emission and absorption of light in atomic sources. These observations 
went on to become the underlying framework for developing the laser.

But the proof of the Planck formula still evaded Einstein. In 1921 he got his
Nobel prize. But the really eventful year for the story we are pursuing was
1924. It was in this year that Louis-Victor-Pierre-Raymond, 7th duc de Broglie
submitted a thesis to the French Academy for a doctorate degree. In it he
proposed that if as per Einstein, electromagnetic waves have a particle like
character, conversely the electron must have a wave character. He proposed the
equally preposterous formula associating a wave of wavelength $\lambda$ with
an electron of momentum $p$.
\begin{equation}
  \lambda = \frac{h}{p}
\end{equation}
This is analogous to the relation $\lambda = c / \nu$ for electromagnetic
waves if we recognise the Planck relation $E = h \nu$, and the Special
Relativistic relation \ for photons, $p = E / c$. It appears that the members of
the Academy were flummoxed by this hypothesis, and after some discussion sent
it off outside France, to Albert Einstein himself for examination. Even
Einstein must have been suitably puzzled. However, he had received the letter
from Bose just a few months earlier. He had now been fully convinced of his
hypothesis of waves behaving as quanta. Much to the Academy's surprise,
Einstein approved De Broglie's thesis proposing material particles behaving
like waves. I would now like to refer back to subsec. \ref{subsec:heuristic}  of part I.
%subsec 4.1
There we considered the possibility  that the reason why Einstein could
withstand the pressure from the stalwarts to withdraw his light quantum paper
was perhaps the fact that the unity of the core concepts for matter and
radiation was more important to him than reconciling the two alternative descriptions
of radiation. The reason why he would readily accept de Broglie hypothesis can
be ascribed to this line of thinking. Until specific writings or records can
be uncovered to support this, it is a matter of conjecture whether Einstein's
ready acceptance of de Broglie's thesis had anything to do with him having
seen a closure to his 1905 hypothesis in the paper of Bose.

de Broglie had an illustrious career as a philosopher scientist. To begin with he was a
duke by inheritance. He quickly became a member of the French Academy. 
His thesis of 1924 proposed that electrons have waves associated with them, which
he called pilot waves. 
%In his 1924 thesis he meant the waves associated with the particles to be in addition
%to the particle itself. 
The waves were supposed to escort the particle like pilot
vehicles in front of the car of a dignitary. de Broglie had also conjectured
that ``the electron has an internal clock that constitutes part of the
mechanism by which a pilot wave guides a particle''\footnote{Wikipedia page on
Louis de Broglie.}. With the development of quantum mechanics, and detailed
consideration of its implications, it has been recognised that there are no waves that
pilot the particle. The particle description and the wave description
are complementary to each other, and mutually exclusive. They are not valid simultaneously. 
This contradicted the original metaphysical motivations of the wave hypothesis. As such 
de Broglie remained vehemently opposed to the subsequent development of his wave ideas 
into wave mechanics. 
% which is founded on the wave and particle descriptions as dual to each other, not to be
%applied simultaneously
Unlike Bose about whose contribution questions
continued to be asked, de Broglie was awarded the Nobel prize in 1929.

To the credit of de Broglie hypothesis is the fact that in the hands of Erwin
Schr{\"o}dinger it bloomed into the landmark new mathematical
formulation of Quantum Mechanics in 1926. Equally importantly, in 1927 the results of the
Davisson and Germer experiments at Bell Labs, scattering of slow electrons
from crystalline Nickel target matched de Broglie's wavelength formula
remarkably.  As for the development of
Quantum Mechanics, Heisenberg was the first in making the radical proposal
that one must abandon the notion of a trajectory in Quantum Mechanics, and
went on to propose the principles of matrix mechanics. In 1925, this version
of the theory was difficult to digest by many as matrices were foreign to
physicists, and the palpable picture that waves offered, and in terms of which
Schr{\"o}dinger's 1926 theory was formulated, rapidly gained acceptance.
Although both formulations are equivalent, the wave formulation holds sway in
most of non-relativistic problems of quantum mechanics. This is somewhat
unfortunate as there are no ``waves'' in the ordinary sense of waves in water
or strings, but only a method to implement the Principle of Linear
Superposition as we explain later.

%\tmtextsf{
\subsection{States and quanta : the essence of quantum physics}\label{sec:StatesandQuanta}
%}

We finally explain the core new conceptualisation of nature that the
Bose-Einstein statistics offers to us. The novel counting that enabled Bose to arrive 
correctly at the Planck-Einstein formula was that in his counting, the states containing 
several quanta received equal weightage regardless of how they were assembled 
from states of single quanta. To understand this, we consider the example of
two coins. Suppose we have two identical coins. And we toss them both
independently. Now we try to anticipate the number of times the various possible
configurations will show up. There are only three possibilities, both heads,
both tails and a third possibility, one head and one tail. We may list these
as HH, TT and HT. Since the coins are indistinguishable, TH is same as HT and
count as the same configuration. However we know very well from classical
experience that out of the total number of possible configurations, the HT (or
TH) is going to occur in two different ways, and hence twice as often compared
to the HH and TT configurations each. The weightage we associate in Boltzmann
statistics with the states of such a system are indicated in the second column
of Table \ref{tab:statistics} under the heading Classical.

\begin{table}[htp]
%\tmfloat{h}{small}{table}
\begin{center}
 \begin{tabular}{cccc}

%\begin{tabular}{|c|c|c|c|}
%& Classical & Quantum:B-E & Quantum:F-D\\
& Classical & Q:B-E & Q:F-D\\
\hline
&&&\\
        H H & $\frac{1}{4}$ & $\frac{1}{3}$ & $0$\\
        &&&\\
        HT or TH & $\frac{1}{2}$ & $\frac{1}{3}$ & $1$\\
        &&&\\
        TT & $\frac{1}{4}$ & $\frac{1}{3}$ & $0$
\end{tabular}
\end{center}
%\label{tab:statistics}
\caption{The weightage factors associated with
  possible states of two indistinguishable coins in Classical, Bose-Einstein
  and Fermi-Dirac statistics.}
\label{tab:statistics}
\end{table}%

However, the Bose-Einstein case is different in a very subtle way. In the
quantum Bose-Einstein counting, the coins are so completely identical that we
are not able to assign twice as much weightage to the HT situation. It has
exactly the same weightage as the HH and TT situations. This is indicated in
the table in the column with heading $Q:B-E$.
%$Q_{B - E}$. 
This kind of counting
applies to particles of spin values in integer multiples of $\hbar$, i.e.,
$0$, $\hbar$, $2 \hbar$, etc. For completeness we have included the last
column which corresponds to Fermi-Dirac statistics obeyed by all particles of
half-integral spin, ie values $\hbar / 2$, $3 \hbar / 2$, .. etc. examples of
which are the electron, the proton and the neutron. If such a species has no
quantum numbers other than the one distinguishing H from T, then the Pauli
exclusion principle forbids HH as well TT. There is only one state admissible
as per quantum principles, HT, with weightage unity!

This state of affairs is called quantum ``indistinguishability'' versus
classical indistinguishability. But as we shall argue, the label of 
"indistinguishability" is predicated on a classical prejudice. And this has 
resulted in enduring confusion and also false hopes of somehow circumventing 
the unpalatable non-classical content of Quantum Mechanics! The suggestion in 
the adjective ``indistinguishable'' is that there are two distinct entities to begin with. 
The quantum logic is however taciturn, and less revealing of its secrets. 
Let us assign a value $+ 1$ to H and $- 1$ to T in
some units. Now the quantum logic allows an observable called the ``number'',
and the value of this number is $2$ in this example, as we have considered two
quanta. However, this system has only one unique state corresponding to total
value $0$ of the H/T quantum number. The availability of the observable
``number'' whose value is 2, should not be confused with there being ``two particles'' in the
classical sense. So the question of ``distinguishing'' between them does not
arise. There is only
{\tmem{one}} quantum state containing {\tmem{two}} particles, the
H/T quantum number of the state being zero, and the weightage of this state is 
exactly the same as that of the other states which have H/T quantum number $+2$ or $- 2$.

At the heart of Quantum Mechanics is the principle of linear superposition.
What we consider classically to be distinct configurations can be ``added'' in
a precise mathematical sense in quantum mechanics. And the weightages of the
superposed states have to be such that the sum of their squares must add up to
unity. As we descend from the macroscopic level to the microscopic, there are
two ways that the quantum rules set in. One is as the systems become simple,
such as small molecules and atoms, systems which are described by a small
number of observables. The other important way quantum principles manifest 
themselves is through the Bose-Einstein or Fermi-Dirac enumeration of states.
But the transition to the fully quantum domain most often is not sharply defined. 
For example, at standard room temperature and pressure, the molecules of hydrogen 
or carbon dioxide gas  obey quantum rules of emission and absorption of radiation, 
but as a collective  system they obey Maxwell-Boltzmann or classical
statistics. What this means is that these behave as independent quantum
systems. For each molecule, its internal states would be a superposition of its
various standard states (in technical language, eigenstates) but the collective
state is not found to be a superposition of some standard states, but rather just like
the states of a small macroscopic particle. This happens because
the gas is very dilute, viz., the average separation between the molecules is about 
50 times larger than their intrinsic size. Typically the collective states of a system 
display quantum mechanical superposition only when the system is densely 
populated with quanta of a given species.

In the domain where the quantum rules apply, the counting of the possible
states becomes different and defies classical common sense. For a variety of
systems even when dense, an approximate picture which allows thinking in terms
of the original isolated quanta works, especially when the quanta have weak
mutual interaction. In such situations one constructs the general states of
the system as products of single particle states. This is purely a
mathematical convenience. Unfortunately this leaves behind the feeling that
two distinct quanta have been put together, even though the symmetrisation or
anti-symmetrisation are applied correctly. Quite a few paradoxes arise simply
from this naive thinking. The ``indistinguishability'' of quantum systems,
more correctly, the appropriate statistics has to be treated as an integral
part of the Quantum principles and not as an added rule. When this is done
consistently, no paradoxes remain, though the rules may continue to intrigue us.

%\tmtextsf{\tmtextbf{
\section{Conclusion and outlook}
%}}

%\tmtextsf{
\subsection{Final story of light}
%}

The novel description of light that began with Planck's formula and was
properly recognised as quantum behaviour of light by Einstein, reached
maturity with the development of Bose-Einstein statistics. While much of the
attention got diverted to condensed matter and nuclear physics, developments
in optics continued separately. One of these was the inelastic scattering of
light due to internal structure of molecules and crystals. This effect,
discovered by \ C. V. Raman earned him the 1930 Nobel prize. While Quantum
Electrodynamics, as a dynamical theory of photons and electrons led to
profound developments, the physics of photons by themselves had entered
quantum era in the 1950's when the maser was developed, soon leading to the
invention of the laser.

In the late 1950's, in the course of using photomultiplier tubes for the study
of stars, Hanbury Brown and Twiss developed intensity interferometry, whose
quantum principles at first seemed to be unclear. By 1963 Glauber and
independently E. C. G. Sudarshan explicated the formalism that applied to the
quantised Maxwell field in all possible settings. Glauber received the Nobel
prize for this development in 2005. In the treatment given by Sudarshan it was
emphasised that the quantum mechanical formula given there accounts for all 
the possible states of light, which subsume the classical states. Put another 
way, what we usually think of as classical Maxwell waves is actually a state of 
the quantised Maxwell field, a special state of the photons. Here the classical 
description applies exactly and no modifications are needed when quantum mechanics 
is taken into account. This formulation  of Sudarshan can be considered to be 
the final closure on the theories of light  originating with Newton and Huygens, 
evolved through the historical path of Planck, Einstein and Bose.

It is intriguing to note that photons have two very special properties, one is zero rest mass 
and the second is zero charge, (or the absence of mutual interactions).  These properties 
have provided us entries, respectively, into the realms of special relativity and quantum 
mechanics.  The zero mass property means that they are always moving at the largest 
limiting speed permissible in nature, ``the speed of light'', and this property has thus 
provided us  a key to Special Relativity.  On the other hand, zero mass and zero charge 
properties have facilitated the observation of the peculiar properties  of a quantum gas 
that we are discussing here. Masslessness means that there is no intrinsic ``size'' to a 
photon such as the Compton wavelength for massive particles. Thus there is no limiting 
dilution in which this gas begins to obey Maxwell-Boltzmann statistics, unlike in the case 
of molecular gases as we noted in sec. 2.2. And the absence of interactions ensures 
that it remains a gas of free quanta to which Bose-Einstein Statistics can be applied in all 
laboratory situations. Massive particles such as atoms also display  quantum superposition 
and enter a collective state called a Bose-Einstein condensate. But to see  this we need 
to prepare their collections with extreme care.  Providing sufficient density may trigger 
interactions; instead, extremely low temperatures are used. For photons, the quantum 
characteristics are readily visible because they constitute a non-interacting quantum gas, 
which enabled the revolution in the hands of Planck.

The theorem of Sudarshan shows that photons provide one more access to the quantum 
world. Dirac has emphasised in his textbook 
"Principles of Quantum Mechanics", that the new content of Quantum Mechanics is the 
principle of linear superposition.  In Classical Mechanics, there is no meaning to a 
plus sign between two possible trajectories of a particle. It cannot be following both. 
In Quantum Mechanics, it is valid to superpose via a plus sign  two states of the system 
which yields a new possible state of the system.  The above theorem
of Sudarshan then has another intriguing implication. Recall that the classical states of 
radiation appear without modification in the complete quantum description. As such, 
the linear superposition principle of electromagnetic fields that is taught at undergraduate 
level is actually nothing but the linear superposition principle of Quantum Mechanics!

%\tmtextsf{\tmtextbf{
\subsection{The enigmatic Quantum}
%}}

Very soon after the basic rules of the new Quantum Mechanics were understood,
it became apparent that the outcomes of experiments could be predicted only
statistically or on the average. Heisenberg had proposed his matrix mechanics
with a clear call that the notion of trajectories must be abandoned. He then
backed up his abstract formalism operationally through a thought experiment,
by showing that attempts to measure one property of the trajectory, say the
position, would necessarily mess up the complementary property, the momentum.
This consequence was natural because the measuring technique itself had to
rely on sending one quantum system, a photon, to ``view'' another, the
electron. It was impossible to improvise any apparatus that was capable of
yielding information of the quantum domain without at the same time obeying
quantum principles. Ergo, it was impossible to beat the uncertainty in
measuring the attributes of a trajectory below a limit set by the new constant
of nature, $h$, or in modern usage, the quantity $\hbar = h / 2 \pi$.

Such a probabilistic outcome given by a fundamental framework was an anathema
to Einstein and to many others of that generation. Needless to say, the
debates continue to rage and are also current. Further, there was the puzzling
property that a quantum state was intrinsically non-local; the wavefunction
was always spread over a space like domain. This seemed to intuitively
contradict Special Theory of Relativity. The enigma of this situation was
formulated by Einstein and his collaborators Podolsky and Rosen with
characteristic clarity and has come to be called the EPR paradox. From a
pragmatic point of view, paradoxical the situation is, but inconsistent it is
not; and no attempts at arriving at an inconsistency \ with the basic tenets
of Special Relativity, even in thought experiment, have succeeded, nor has a
clever experiment been designed that would force an extension of Quantum
Mechanics.

The other puzzling aspect of Quantum Observation is that only specific
eigenvalues are returned as the outcome of measurement. For instance the
average spin of an electron in a beam may be $0.35 \hbar$, but that only means
that if you made measurements on many electrons in that beam, that would be
the average outcome. In any one specific measurement that manages to capture
only one electron, the answer will be precisely either $+ \hbar / 2$ or $-
\hbar / 2$.

This fact has been well verified. But it leads to the following paradoxical
situation called Wigner's Friend or Schr{\"o}dinger Cat depending on how
amicable or macabre your inclination is\footnote{Actually Wigner's Friend is a paradox
which is a step beyond the more direct paradox presented by Schr{\"o}dinger's Cat. Due to brevity
of the presentation here, and presuming that many readers are already familiar with these paradoxes,
I have taken the liberty to speak of them together.}. Once measured, the system will go on
being in that eigenstate, say spin $+ \hbar / 2$. But now if you sit quietly
after that measurement, your friend who walks in has no way to decide whether
it is already in an eigenstate or not without actually making the measurement
herself.  The dilemma at hand can be seen to result from the rule of quantum mechanics, 
that once an attribute is measured, the net effect of the measurement 
process is to leave the system in one of the eigenstates of the particular observable. 
But this rule creates an unequal situation for different categories of
observers, those who first carried out such an observation on a generically prepared system, and those
that come later, without knowing whether the measurement has been made by some other party.  
Thus the well established notion objectivity even in the classical world of observers seems to be endangered.

Nobody wants to kill a cat ever, let alone twice, so such a paradox jumps out
to challenge common sense. But the resolution is very simple. If the first
observer has already made the measurement, then the system can be considered
to have been prepared in that specific eigenstate for the next observer. No
contradictions arise but the challenge to common sense persists. Also one
hopes that a refinement of the formalism would make measurement a more
intrinsic and organic part of the formalism than a drastic ``collapse'' into
specific eigenstates. Constructing such a formalism is an active area of research. 
But we would expect that such a \ formalism will only be an extension, without modifying 
any of the core tenets of Quantum Mechanics.

Since the Quantum is maligned so much in common discourse, it is worth
emphasising that there is a lot that is counter intuitive in Physics. As we
know, understanding the phenomenology of classical motion was itself a great
intellectual enterprise, culminating with the discourses published by Galileo.
Its final refined version we accept with equanimity is due to Newton. Yet,
there is much that is conceptually unsatisfactory about Newtonian framework which we have come
to take for granted. The foremost among them is the notion of limits as needed
for the infinitesimal calculus. Through the formal concept of instantaneous
velocity, we are convinced by Newton that a particle can be at a point and
also moving while still being at that point! In fact it is supposed to possess
all orders of time derivatives while still just being at its original point. 
In the bygone era of theology this would have remained an active area of
debate, but not so in modern engineering. While the high level of refinements
in real analysis ensure that there is no logical contradiction, the point
remains that this is a mind game. Operationally it is impossible to make your
stop watch measure vanishingly small duration. Indeed, now we know that
Quantum Mechanics will kick in and will show that the Newtonian process of a
limit is a figment of our imagination. 

It is time we accepted that our intuition is based on the cognitive faculties
tuned to the classical experience. And that physical science requires a kind
of sophistication that may yield counter-intuitive theorems. And some of the
puzzles will fade from common discourse, much as theology of yester years, as
highschool students begin to interact with quantum systems and the quantum
framework earlier in their physics syllabus.
\section*{
%\tmtextsf{
References}
%}
\begin{enumeratenumeric}
  \item Wikipedia articles for various specifics of dates, personalities and
  experiments
 
  \item ``Einstein's Miraculous Year: Five Papers that Changed the Face of
  Physics'', edited and introduced by John J. Stachel, Princeton University
  Press, 1998  
 
  \item ``Subtle is the Lord : The Science and the Life of Albert Einstein'',
  Abraham Pais, Oxford University Press, 1982
 
  \item ``Inward Bound: Of Matter and Forces in the Physical World'', Abraham
  Pais, Oxford University Press, 1988
\end{enumeratenumeric}

\end{document}